# Using LLMs to Evaluate Architecture Documents – Results from a Digital Marketplace Environment

Frank Elberzhager [0000-0002-8748-3927], Matthias Gerbershagen, and Joshua Ginkel

[1] Fraunhofer IESE, Fraunhofer Platz 1, Germany
`{first.last}@iese.fraunhofer.de`

**Abstract.** Generative AI plays an increasing role during software engineering activities to make them, e.g., more efficient or provide better quality. However, it is often unclear how much benefit LLMs really provide. We concentrate on software architects and investigated how an LLM-supported evaluation of architecture documents can support software architects to improve such artefacts. In the context of a research project where a digital marketplace is developed and digital solutions should be analyzed, we used different LLMs to analyze the quality of architecture documents and compared the results with evaluations from software architects. We found out that the quality of the artifact has a strong influence on the quality of the LLM, i.e., the better the quality of the architecture document was, the more consistent were the LLM-based evaluation and the human expert evaluation. While using LLMs in this architecture task is promising, our results showed inconsistencies that need further analyses before generalizing them.

## 1      Introduction

Software development is at a turning point. A survey by Daigle et al. showed that almost all software developers stated that they had already tried AI coding tools, and about 60% expect them to improve the fulfillment of customer requirements [1]. However, even the most powerful AI systems are subject to a fundamental principle: the quality of their output is largely determined by the quality of the input. While humans implicitly use domain and company knowledge, AI systems only use the knowledge acquired during training and explicit input. This makes documents that are developed early in the software development process, such as specifications or architecture, and which form the basis for later implementation, a critical success factor for dealing with AI. It is important to ensure the best possible quality of these software artefacts, as issues, defects, ambiguities, missing information, and inaccuracies early in the development process lead to potentially expensive correction cycles if they are implemented and then only found late in testing activities or even by the user.

During software development, there are therefore a variety of methods that support early quality assurance, which is reflected both in constructive support (for example in the form of design and architecture patterns) and analytical support (for example in the form of reviews or static model analyses). Based on the observation that implementation is increasingly being carried out autonomously or at least with support of generative AI, it is even more crucial that the quality of the input is maximized. Typically, different architecture diagrams, architecture drivers and textual descriptions are a major input for coding activities. Analyzing the quality of architecture documents is nothing new and many approaches, from experience-based over tool supported to structured analyses, exist [3]. We were interested in how generative AI can be used in those early software development phases to support architecture tasks.

Before we strive towards higher automation in creating architecture artifacts, for example by using agents that develop certain architecture models, we first wanted to analyze whether generative AI can evaluate different architecture documents and models. We wanted to understand whether generative AI provides adequate results in judging the resulting quality of architecture documents and how this can be compared to architecture experts. As there is usually not "the one" architecture and there is even discussions among software architects, it is relevant to understand how the AI performs and whether this can become a real support (and, for example, can make software development more efficient or of higher quality) or whether the results are more anecdotal. Therefore, the focus in this article and our main research question is:

- Under the assumption that generative AI can be used to evaluate architecture documents, what is the quality of the evaluation compared to human software architects?

In order to start answering this research question, we analyzed software repositories from a digital marketplace[1]. While we develop and operationalize the marketplace, several solution providers can offer their digital solutions, apps and services on that marketplace. The focus is on solutions in the area of smart cities. Municipalities can decide which digital solutions they would like to use and offer their citizens. Because of many solutions provided on the marketplace, municipalities asked for some kind of quality evaluation of the solutions, and defined a set of quality criteria, such as documentation quality, community aspects or architecture. Such criteria are then automatically checked by an AI-based component, and we were able to draw first conclusions about the general applicability of generative AI for analyzing different artifacts. As such quality criteria are, however, typically on a higher abstraction level than what we would think of regarding a detailed evaluation of architecture documents, we performed an additional and more fine-grained analysis of available documents in order to understand better how good generative AI can do such a task.

The remainder of the paper is structured as follows: Section 2 provides related work regarding how generative AI is used in architecture work. Section 3 shows some details about the marketplace in order to understand better our evaluation setting. Section 4 then describes our solution that is able to perform the automated analysis of different input documents with generative AI. Section 5 provides details about our evaluation

---

[1] https://marktplatz.deutschlanddigital.org/

and discusses the results, and Section 6 summarizes the paper and gives an outlook on next steps.

## 2    Related Work

### 2.1   Quality Evaluation in Software Architecture and Software Architecture Documentation

As Bass et al. [2] have previously outlined, there are no universally applicable guidelines for good software architectures, as these are always tailored to meet the specific requirements of the project. Consequently, architecture can be considered either suitable or unsuitable, but not necessarily good or bad. To ensure a high standard of quality, several methods and procedures have been established over time.

One of the first established ones is the Architecture Tradeoff Analysis Method (ATAM). The process is divided into four phases. First, teams are assembled, then information is collected. Next, the information is analyzed. Then, risks, quality requirements, and important decisions are documented, and finally presented. The participating groups consist of an external evaluation team, the project decision-makers, and the architecture stakeholders.

The Software Architecture Analysis Method (SAAM) is a systematic approach consisting of five phases. During the first phase, the system is divided into basic functions. Following, these basic functions are mapped to the architecture. The third phase involves specifying quality characteristics and defining benchmarks. The result is an assessment of the degree of fulfilment and difficulties [2].

An approach specifically designed for the evaluation of architecture-related documents can be found in the Documentation Quality Check (DQC) within the framework of Fraunhofer RATE (Rapid ArchiTecture Evaluation). At the outset, all artifacts are collected and knowledge about the target groups of the architecture documentation is gathered. Using surveys or perspective-based reading [4], a model is created for each target group, for example, an instruction manual for the users of the product. An evaluation is based on whether all relevant information was included in the creation process [3].

Furthermore, there are various approaches and tools that combine both manual and automated processes to ensure the quality of software architecture documentation. A common manual approach is to use best practice checklists, which architects use to systematically check documentation for completeness, comprehensibility, and consistency. These checklists are often based on established quality criteria and serve as a tool for identifying typical errors and ambiguities at an early stage.

Beyond purely manual procedures, several automated approaches have been developed that analyze documents for quality aspects. The Automatic Checking of Quality Best Practices in Software Development Documents tool checks development documents against around 80 rules relating to naming conventions, duplicate detection, and word choice, among other things [5]. The aim is to automatically detect recurring quality defects and thus reduce the effort required for manual checks.

Another example is the Document Quality Checking Tool for Global Software Development, developed by IBM Japan in 2012. This tool was designed specifically for the challenges faced by globally distributed development teams, where differences in time zones, languages, and cultures often lead to inconsistencies in documentation. It uses a comprehensive set of 144 checking rules to identify linguistic and structural quality issues and ensure a consistently high level of documentation quality [7].

In addition to these documentation-specific tools, general automation technologies also contribute to quality assurance. These include the automation of testing, deployment, and debugging, as well as the translation of outdated code into modern programming languages and the generation of test data and are able to improve the efficiency [6].

### 2.2  LLMs used for software engineering and software architecture

Generative AI also plays a strongly growing role in software engineering activities and considers all phases, from requirements engineering to architecture to implementation and testing. Hou et al. [11] provide a large literature review on large languages models for software engineering and sort literature to six phases, with a focus of identified papers for software development, quality assurance and maintenance. Only a very limited number of papers were found to support the design phase. In general, there was an explosion of papers from the year 2022 to 2023, and this trend continued.

Jahic and Sami [12] provide a state-of-the-practice analysis on LLMs in software engineering and software architecture. They conducted an interview-study and performed LLM-assisted architecture design. The authors claimed potential in these activities but also stated reasons that have to be overcome in the future such as copyright issues, low quality and checking efforts of the LLM-based results.

Schmid et al. [13] performed a literature review analyzing which tasks can be supported by LLMs. Based on 18 papers identified, typical tasks such as generating code from architecture models, analyzing or supporting decision making are not a focus until now, but get a growing interest.

It can be concluded that there is a huge and even growing number of papers on almost all relevant software engineering conferences, journals and magazines nowadays and that genAI will play a growing role in software engineering tasks.

## 3  Use Case and Implementation Insights

To investigate the potential of generative AI when evaluating software architecture documentation, we started our research in the following project setting: Evaluation of the architecture documentation of open-source solutions in the smart city context. The output of our tool should give helpful support for municipalities concerning the quality of the architecture documentation so they can decide on the use of an open-source solution. The results of the evaluation are integrated into "Deutschland Digital", a marketplace that has been developed for municipalities. We used the evaluation criteria provided by the German Federal Ministry for Housing, Urban Development and Building. In the domain of software architecture, the main criteria to evaluate are modularity, configurability, code quality and documentation.

For our first prototype we chose a small set of open-source solutions and did the evaluation with the help of LLMs and a static code analysis tool (Sonarqube[2]). We tried to evaluate all criteria given by the ministry to cover a broad range of aspects. Most of the evaluation was done by using an LLM. Due to the small set of solutions under evaluation, it was easy to just use an LLM for almost everything and tune the prompts accordingly. With this approach we covered most aspects and got early adequate results. Of course, such a solution does not scale very well, and the results are hard to reproduce.

To provide meaningful decision support for municipalities, the results of our evaluation needed to be more stable over time. Furthermore, the service needs to be able to evaluate all open-source solutions provided in the marketplace. Currently, there are more than 100 digital solutions, and the number is increasing by about two to five solutions per week. This means we needed to consider scalability of the service and costs for LLM calls. Consequently, for the production-ready service we choose a different approach than for the initial prototype. Most criteria are evaluated using deterministic code and established tools, like for example scorecard [9]. LLM-based evaluations are only done for a handful of sub-criteria, which are hard to evaluate using a deterministic approach, like for example judging whether the language used in the issues in the issue tracker of the project is rude or not. In the final service, the LLM is just another tool in our toolbox. But still some criteria of the prototype are not evaluated in the final service at all, because to evaluate them, we would require additional context information, for example about the execution of certain organizational processes in the municipality. Beyond that, we choose to use weighted quality models [10] to aggregate and structure our evaluation in a deterministic way. Our quality models used for the evaluation are stored in our database and are not part of the source code. This choice also allows us to dynamically adjust the evaluation (for example, to change the weight of certain sub-criteria) without changing the code of our service. In Figure 1, the aggregated results of such a quality- model based evaluation in the marketplace can be seen.

---

[2] https://www.sonarsource.com/products/sonarqube/

## 4  Automated Evaluation of Documents with LLMs

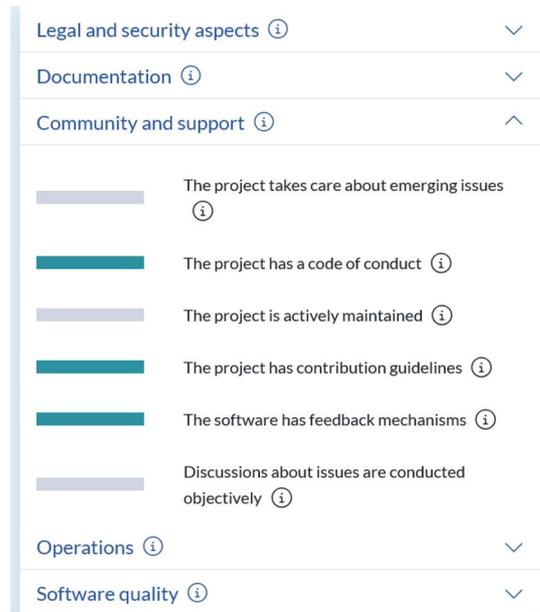

*Figure 1: Excerpt of an Evaluation of a Marketplace Solution*

We have developed an MVP of an LLM-based tool called "Quasar" to support the evaluation of software quality. It uses different software engineering (SE) artifacts as input and generates quality scores as output. These scores are used to evaluate different qualities of software solutions. Figure 2 illustrates the essential building blocks of our solution idea in the marketplace setting.

For each relevant aspect (e.g. architecture documentation, operations, or software qualities such as usability in particular), we have developed a quality model that supports a systematic evaluation. Quality models combine and weight measurements of different metrics to assess quality in a systematic way [10]. The model defines threshold values that determine whether the desired quality level has been achieved or not. Traditionally, measurements related to the evaluation of architectural documentation are performed manually: human experts analyze the available material, talk about it with relevant stakeholders, understand the system context, and then evaluate the documentation quality based on conclusions. We first wanted to evaluate to what extent we could replace the human expert with an LLM-based expert. The goal was to find out whether our solution would be able to assess the quality of the documentation and provide a rationale for the assessment that is plausible.

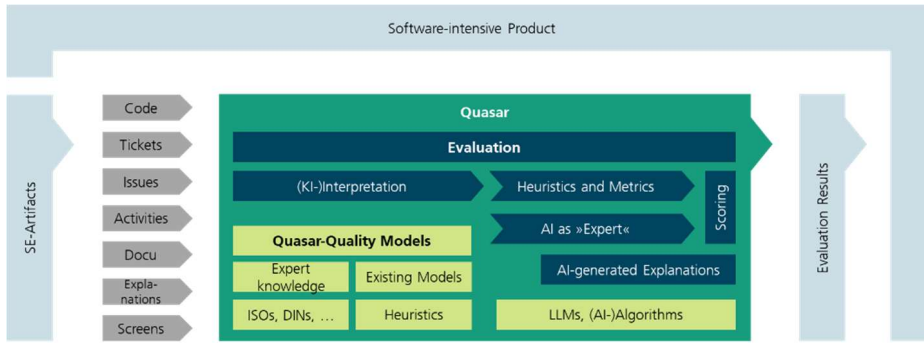

*Figure 2: General Workflow and main Components of the Quasar Solution to Evaluation Architecture Documents from Digital Solutions provided on a Digital Marketplace*

Our solution does not expect a specific architecture documentation structure. However, certain predefined content types are expected (in the case of underlying architecture documentation, for example, an introduction, naming of stakeholders, delimitation of the system context, a functional breakdown, explicit design decisions). Various documentation formats are then initially processed by a converter. This uses an external library that can handle many common documentation formats such as PDF, DOCX, Markdown, etc.

Quasar is divided into two layers: the UI and the backend (see Figure 3). The UI serves as a lightweight way to interact with the MVP through a web browser and to launch requests to analyze development projects. The backend processes the requests, downloads the files of the development project, filters and analyzes suitable files, supplemented by further architecture documentation that can be provided via the UI. Different evaluator components in the backend then take care of analyzing the evaluation criteria. First, it converts data into a format suitable for the LLM using the file or image converter. Then it uses a set of prompts for different LLMs to evaluate the criteria with the help of the LLM client component. As a last step, the evaluation component combines the responses from the LLMs into the evaluation result.

We used different models from different vendors for our prototype. Therefore, we introduced the LLM client component. Its main purpose is to take care of the interaction with different LLMs. With this component, we reduced the coupling of the evaluator components to the concrete LLM. This facilitates subsequent extensions or changes and makes it easy to address other models in the future.

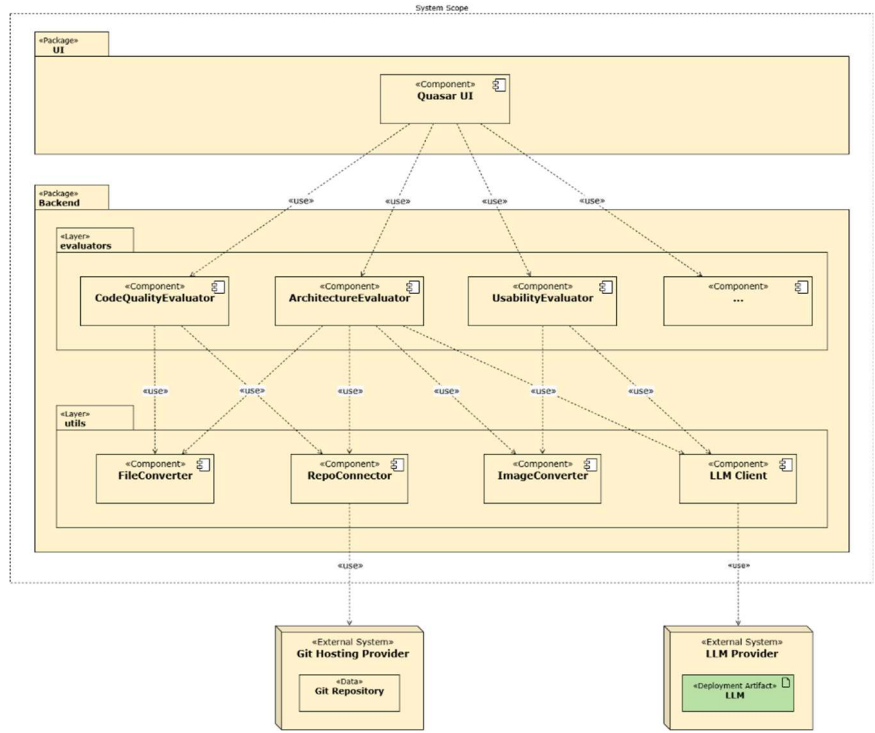

*Figure 3: Architecture of the Quasar Solution*

Another important decision was to divide the evaluators according to use cases. For example, there is an evaluator for architectural documentation, but also for qualities such as usability. The main advantage of this is the simplified parallel development of evaluators by different teams, as well as a clear separation of use cases and domains.

## 5 Evaluation

### 5.1 Pre-evaluation of Architecture Documentation

In connection with our digital marketplace, the analysis of architecture documentation was carried out on open-source development projects that are actively used by users in practice (and the publicly accessible architecture documentation). The marketplace was developed to support the improvement of services of general interest in rural regions. The special feature of the marketplace is its nationwide availability as well as the search and mediation of quality-assured and needs-based digital offers. Municipalities, for example, can find and purchase digital solutions for specific challenges in the marketplace. These solutions can be placed on the marketplace by providers and are available after a review. Quality assurance for a scalable marketplace is not possible (or very time-consuming) through manual evaluations by experts alone, which also includes analyzing the architecture documentation. This is where our Quasar solution supports: an initial, lightweight quality assurance of the submitted solutions. With the help of genAI, an evaluation is now carried out to support experts and save time. An example of such an evaluation is shown in Figure 1 before. There, five criteria can be seen with some more details regarding community and support.

In the use case described, it was sufficient to make a general statement on the quality of the architecture documentation in order to provide a rough orientation and to ensure that such documentation was available at all. However, we realized early on that there are stakeholders who are much more interested in detailed information. Therefore, we shifted more towards software developers and especially software architects to provide information on the quality of the architecture documentation (and other findings). With this, we can directly give feedback to the developers of the solutions on the marketplace so that they can improve their solutions and the documents developed.

### 5.2    Detailed Evaluation of Architecture Documentation

**Goals, metrics and procedure**

The aim of this evaluation is to gain an impression of the comparability between human and AI-based assessments of software architecture artifacts. We wanted to gain more clarity on the question of whether or how LLMs can support experts in their daily work or even take over parts of it. For this study, we selected two solutions from different solution providers on the marketplace and took a closer look at them. One of the projects provided comprehensive documentation of its architecture, while the second project only offered basic information. For an objective investigation, we compared the agreement of humans with that of Quasar on a scale from 0 (disagree completely) to 4 (agree completely) based on 25 statements.

To achieve this, two software architects were surveyed, and Quasar was applied to each project three times. As there is no official, universally valid evaluation to use as a basis for comparison, the individual runs of Quasar are checked for consistency and completeness, and the average human evaluation is compared and examined for similarity. Both used the same evaluation sheet.

An overview of the procedure can be found in Figure 4.

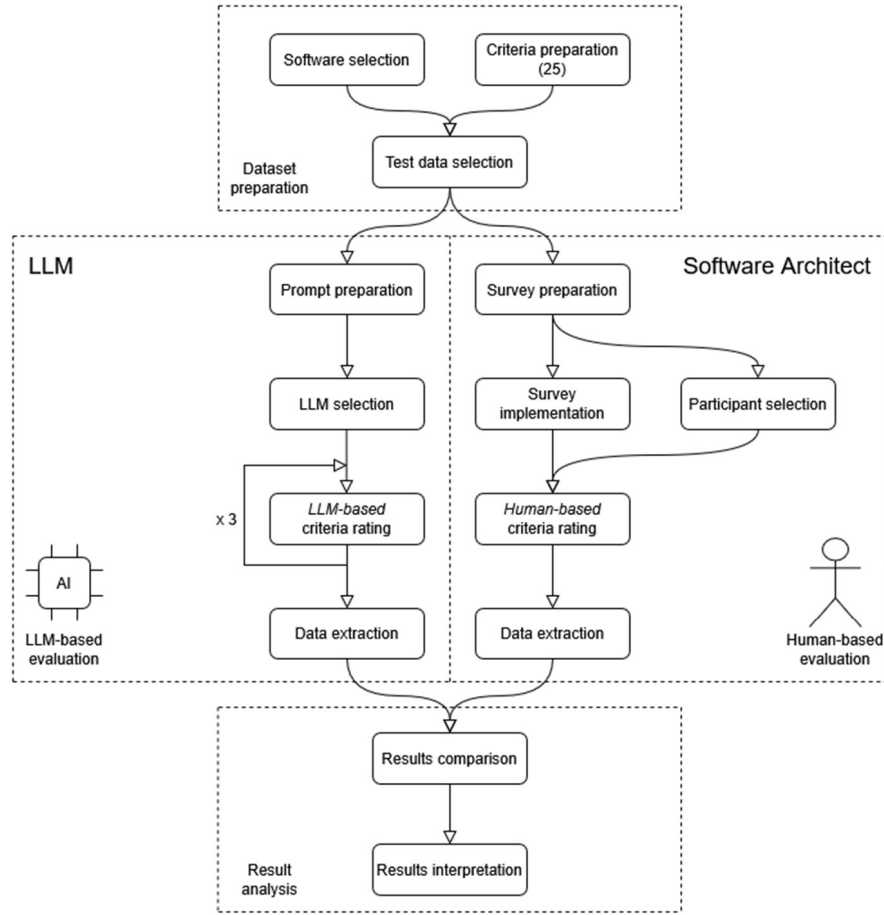

*Figure 4: Evaluation Procedure*

**Dataset preparation**

The basis for evaluating architecture documentation is the collection of all relevant artifacts. To begin with, both the repository of the relevant project and its wiki are downloaded from GitHub. From all files, those that contain potential architecture information are filtered out by first converting the file in question into a readable format using Docling[3] and then comparing its content with a list of keywords. The files found, which include images, diagrams and text, were evaluated using LLMs.

The first project utilizes GitHub Pages to present its content. Its documentation folder comprises a total of 266 files, including 95 Markdown files, 23 PDFs, and 126 images. The term "Architecture" occurs 11 times. The second project employs the GitHub Wiki, which contains only 11 Markdown files with 25 embedded images. The word "Architecture" was found three times. Projects with varying documentation sizes were deliberately selected to investigate how the extent of documentation influences the results.

---
[3] https://www.docling.ai/

**Model selection**

The specific models used were Qwen2[4] and Llama3.1[5], which we host locally on a hardware with three NVIDIA Tesla V100 SXM3 32 GB GPUs. While Llama3.1 only processes text, Qwen2 is a multimodal model that can process both text and image inputs, making it possible to analyze diagrams not only in isolation, but also in conjunction with accompanying text.

**Instrumentation and data collection**

The overall evaluation was based on 25 predefined evaluation criteria. There is a corresponding prompt for each of these criteria. An example of such a prompt is: "The key design decisions relating to the solution concepts are presented and justified." All criteria / prompts can be seen in Appendix A.

A file, including any images, is sent to the LLM together with the respective prompt, which returns a score as an evaluation. Each file is evaluated according to the 25 defined criteria, and the overall evaluation is composed of the respective individual evaluations. If a file does not contain any relevant information, this is considered in the overall evaluation by excluding this file from the evaluation for the specific criterion. If a criterion is not found in any file, this results in a poor evaluation of this criterion. The process is controlled by a Python program. The focus is on the evaluation in the form of a score and less on a textual justification for this evaluation, but Quasar still outputs a machine-readable file that contains a score for each evaluation criterion as well as a summary justification generated by the LLM.

**Research questions**

The overall research question was already stated as whether generative AI can be used to evaluate architecture documents and what is the quality of the evaluation compared to software architects. We refined this question into three sub-questions for a more detailed analysis:

RQ 1.1: How effort-intensive is the evaluation of the architecture documentation by Quasar compared to human architecture experts?

RQ 1.2: How consistent is the result from Quasar if repeated several times?

RQ 1.3: How is the quality of the evaluation from Quasar compared to human architecture experts?

**Results**

The results of the evaluations showed significant differences between the projects.

---

[4] https://huggingface.co/collections/Qwen/qwen2
[5] https://huggingface.co/collections/meta-llama/llama-31

**RQ 1.1 (project 1):** In the first project, the processing time was approximately 60 minutes for humans and 68 minutes for Quasar.

**RQ 1.2 (project 1):** The three automated runs of the model showed a largely consistent evaluation (see Figure 5).

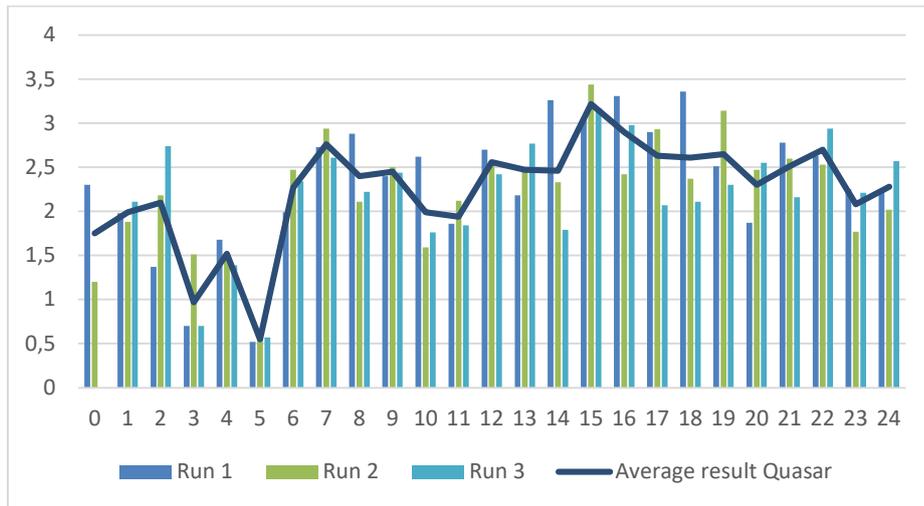

*Figure 5: Overview of the Consistency of the Three Runs (Project 1)*

**RQ 1.3 (project 1):** In comparison to the expert, there were almost exact matches for two statements, while 36% of the assessments differed by a maximum of 15%. The average deviation was 27% (see Figure 6).

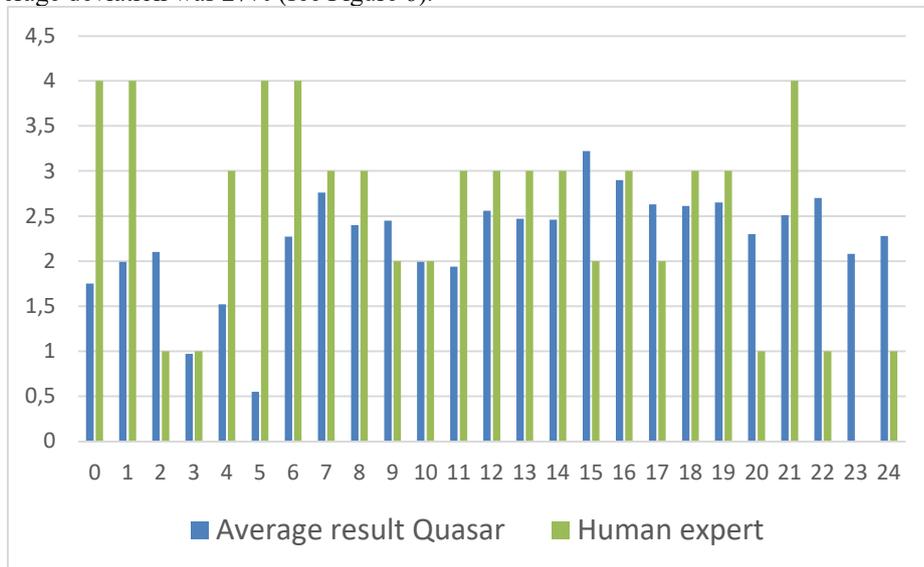

*Figure 6: Comparison of the average Deviation between Quasar and the Human Expert (Project 1)*

**RQ 1.1 (project 2):** In the second project, the processing time was significantly shorter at 22 minutes (human) and 14 minutes (Quasar). In this instance, the model was unable to provide a result for three statements, while the overall average deviation was 43%, implying that there is virtually no discernible apparent correlation.

**RQ 1.2 (project 2):** The three automated runs of the model showed a weak consistency. For three statements, Quasar was unable to produce a result in any run (see Figure 7**Error! Reference source not found.**).

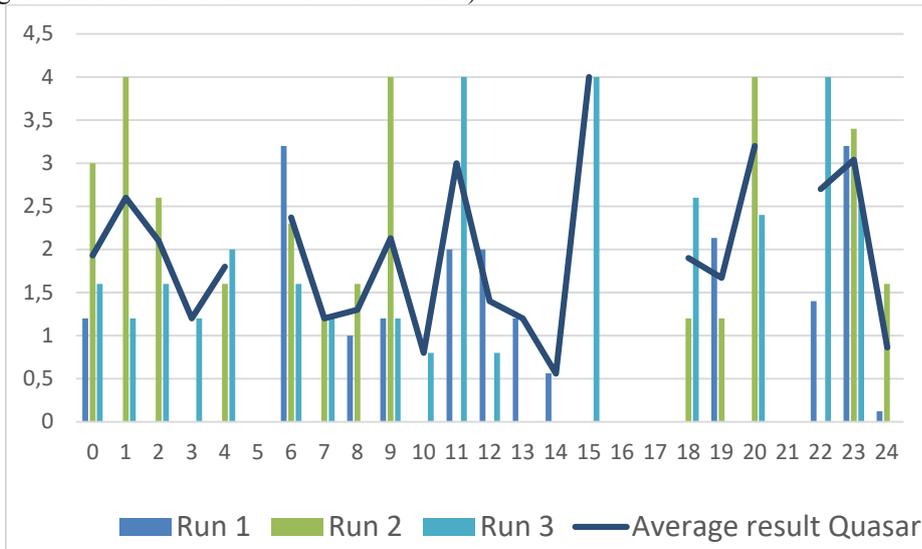

*Figure 7: Overview of the Consistency of the three Runs in Project 2*

**RQ 1.3 (project 2):** In comparison to the expert, no apparent correlation can be established. All evaluations show a significant difference with an average deviation of 43% (see Figure 8Figure 6).

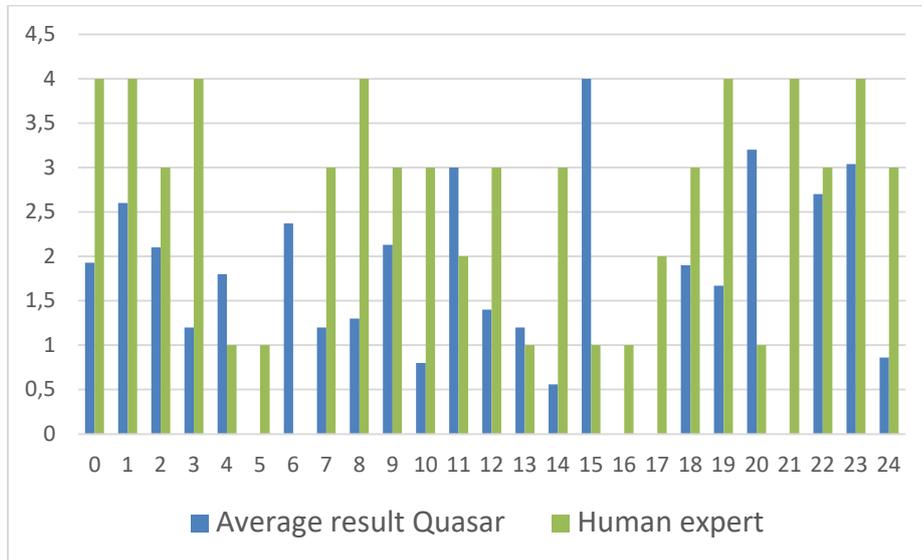

*Figure 8: Comparison of the average Deviation between Quasar and the Human Expert (Project 2)*

**Discussion and Threats to Validity**

The results of the evaluation show both the potential and the limitations of using LLMs to evaluate architectural documentation. We were able to show that AI models are capable of reliably recognizing simple quality aspects and evaluating them in a structured manner based on statements, while more complex relationships can only be captured to a limited extent at this stage.

By observing that the quality of the input material has a decisive influence on the quality of the LLM outputs, we were able to deepen our assumptions. The more structured and complete the artifacts were, the stronger were the apparent correlation between the LLM evaluations and the assessments of the software architects. In contrast, when the information density was low, there were significant deviations.

From a technical perspective, there have been various innovations since the study was conducted, which are described in more detail in Section 5.3. Due to the use of smaller models and the decision not to use retrieval-augmented generation, we had to develop our own data management and filtering system, which sorted sections in or out using a list of keywords. This means that some information may have been lost.

The number of projects that we considered, and the number of human experts, was rather low so that generalization is not possible based on the current results. We also had to rely on the accuracy of the experts' assessments. These could never be completely objective, as every person is influenced by their abilities and experiences.

Regarding RQ 1.1, the results fully met our expectations. A larger amount of data requires more time, while a smaller amount requires less. This was reflected both in machine processing and in human evaluation. The time effort between humans and the machine was very similar in our assessment. However, an assumption for future evaluation is that if the document size increases, the automated solution will be faster. In this case, however, quality will also play a major role, i.e. to evaluate whether saved time comes with a drawback in quality or whether the genAI solution provides adequate results one can rely on. Many further detailed questions are reasonable, of course.

For RQ 1.2, an apparent correlation could be observed between the quality and volume of the input data and the consistency of the results. Higher-quality input reduces hallucinations and allows the task to be executed in a more structured manner.

The same relationship is evident for RQ 1.3. With high-quality input, the evaluations of humans and Quasar deviate only slightly, whereas insufficient input leads to greater discrepancies.

Overall, the study shows that LLMs are a useful support for human work and can provide an initial impression of the quality of software documentation, which is a valuable addition, especially for the marketplace use case.

### 5.3   Evaluation Optimizations for Future Iterations

For future evaluations, several optimizations are possible to obtain a more defined impression of the comparability between human experts and machines.

Firstly, the number of participants to date has been too small to allow for universally valid conclusions to be drawn. The expertise and assessments of individual people are heavily dependent on their experience and education, making it necessary to calculate an average from a larger number of participants. The same applies to the number of projects and (architectural) documentation examined.

From a technical perspective, there have been numerous developments that simplify data processing. In the meantime, the models used have already been replaced, and a switch to a cloud provider was made to achieve better stability and scalability. Current versions of Google's Gemini, for example, offer input lengths of over one million tokens [8], which would be sufficient to process most documentation in its entirety. Another option would be to use Retrieval-Augmented Generation (RAG) to store information in a structured manner and retrieve it as needed. Both variants would guarantee virtually lossless data processing and avoid fragmentation of the input, which would lead to the preservation of contexts across multiple files. By additionally considering the reasoning of humans and LLM, coincidentally similar reasoning could be filtered out, which would strengthen the robustness of the evaluation. Further improvements are hoped for by fine-tuning a model to strengthen domain-specific knowledge.

## 6      Summary and Outlook

In this paper, we presented an evaluation of architecture documents with support of generative AI. We compared the results of our Quasar solutions with human architecture experts and could show that general support can be provided – meaning the solution is a promising way to support human architects. The evaluation was based on a small set of participants and documents, so the generalizability has to be shown with further evaluations in the future. Our tool solution Quasar is built in a way to be flexible, i.e.

further evaluations of quality can be added easily, and new or alternative LLMs can be considered when better language models are developed or further evolve.

In general, using generative AI to support software engineering roles such as architects is a promising way to, e.g., improve efficiency, provide better quality or give new ideas in designing software systems and digital solutions. We therefore also see potential in developing plugins for tools that software architects use in their daily work, such as Enterprise Architect or draw.io. Such plugins can directly provide feedback during design activities and might provide a smoother working environment compared to a separate tool such as Quasar currently is. In addition, besides evaluation tasks performed or at least supported by a generic AI solution, constructive activities might be a future step, especially when thinking into the direction of agents. We will focus on these two directions basically, i.e., more evaluation and further development to better support software architects.

**Acknowledgments.** We thank the software architecture experts for supporting us in the evaluation.

# References

1. Daigle, K. et al., 2025: Survey: The AI wave continues to grow on software development teams. Online abrufbar unter https://github.blog/news-insights/research/survey-ai-wave-grows/.
2. Bass, L., Clements, P., Kazman, R.: Software Architecture in Practice, 3rd edn. Addison-Wesley, Boston (2012)
3. Knodel, S., Naab, M.: Pragmatic Evaluation of Software Architectures. Springer International Publishing, Cham (2016)
4. Basili, V.R., Green, S., Laitenberger, O. et al.: The empirical investigation of Perspective-Based Reading. *Empirical Software Engineering* **1**, 133–164 (1996). https://doi.org/10.1007/BF00368702
5. Dautovic, A., Plösch, R., Saft, M.: Automatic checking of quality best practices in software development documents. In: Proceedings of the 11th International Conference on Quality Software (QSIC 2011), pp. 208–217. IEEE, Los Alamitos (2011)
6. Ebert, C., Louridas, P.: Generative AI for software practitioners. IEEE Software 40(2), 30–38 (2023)
7. Miyamoto, K., Nerome, T., Nakamura, T.: Document quality checking tool for global software development. In: 2012 Annual SRII Global Conference, San Jose, CA, USA. IEEE (2012)
8. Comanici, G., Bieber, E., Schaekermann, M., Pasupat, I., Sachdeva, N., Dhillon, I., Blistein, M., Ram, O., Zhang, D., Rosen, E., Marris, L., Petulla, S., Gaffney, C., Aharoni, A., Lintz, N., et al.: Gemini 2.5: Pushing the Frontier with Advanced Reasoning, Multimodality, Long Context, and Next Generation Agentic Capabilities. arXiv preprint arXiv:2507.06261 (2025)
9. Scorecard HomePage, https://openssf.org/projects/scorecard/, last accessed 2025/10/28
10. Wagner, S., Goeb, A., Heinemann, L., Kläs, M., Lampasona, C., Lochmann, K., Mayr, A., Plösch, R., Seidl, A., Streit, J., Trendowicz, A., "Operationalised product quality models and assessment: The Quamoco approach." Information and Software Technology 62 (2015): 101-123.

11. Hou, X., Zhao, Y., Liu, Y., Yang, Z., Wang, K., Li, L., Luo, X., Lo, D., Grundy, J., Wang, H.: Large language models for software engineering: a systematic literature review. arXiv preprint arXiv:2308.10620 (2024). https://arxiv.org/abs/2308.10620
12. Jahic, J., Sami, A.: State of practice: LLMs in software engineering and software architecture. In: *Proceedings of the ICSA 2024 Workshop on Software Architecture Machine Learning (SAML)*. Springer, Cham (2024)
13. Schmid, L., Hey, T., Armbruster, M., Corallo, S., Fuchß, D., Keim, J., Liu, H., Koziolek, A.: Software architecture meets LLMs: a systematic literature review. *arXiv preprint arXiv:2505.16697* (2025). https://arxiv.org/abs/2505.16697

## Appendix A

Set of questions that were used as prompts in our evaluation of architecture documents.

0. The documentation starts with high level concepts and descriptions (no technical details required).
1. The documentation introduces the basic idea of the system, meaning the system goals are, the motivation for the system existence, the key challenges that the systems should solve, and its business context.
2. In the introductory part of the document, key functional requirements of the system are introduced. (If they are provided explicitly, the rating should be better; if implicitly, less good; if not provided at all, the rating should be 0).
3. In the introductory part of the document, key quality requirements are mentioned. (If they are provided explicitly, the rating should be better; if implicitly, less good; if not provided at all, the rating should be 0).
4. The documentation introduces the system's constraints. (If they are provided explicitly, the rating should be better; if implicitly, less good; if not provided at all, the rating should be 0).
5. The documentation introduces the system stakeholders. (If they are provided explicitly, the rating should be better; if implicitly, less good; if not provided at all, the rating should be 0).
6. A system context delineation diagram is provided and there is textual explanation for the diagram elements.
7. The list of functions exposed by the system to the external actors (be it humans or external systems) is provided.
8. The data exchange between the system and external actors are described. The description names the data that goes in and the data that goes out of the system.
9. A domain model is provided to illustrate the main entities of the domain and their relationships.
10. The documentation provides the meaning of the key entities of the business domain.
11. The documentation describes the main functional requirements of the system.
12. The documentation provides the quality requirements of the system.

13. The quality requirements are described in a way that the desired behavior of the system can be measured. (and therefore verified) when the solution is implemented.
14. The documentation describes the fundamental solution strategy for the system functional requirements.
15. The documentation describes solution concepts for addressing all the system's functional requirements.
16. The documentation describes solution concepts for addressing all the system's quality requirements.
17. The key design decisions related to the solution concepts are presented and justified.
18. The documentation provides a functional decomposition of the system at runtime (i.e., what components and subcomponents the system is composed of).
19. The documentation provides the data flow among different components in the system.
20. The documentation describes the deployment strategy of the system. It includes a deployment diagram. It contains typical elements of a deployment diagram such as the deployment artifacts, the execution environments, the computing nodes, storage systems, etc.
21. The documentation provides information about the key technologies used to implement the system (e.g., programing languages, storage system, libraries, frameworks, etc.).
22. The documentation describes the modules (development time elements) that implement the runtime components.
23. The documentation describes the needed tools and steps to deploy the system.
24. The documentation describes the data model of the system (e.g., class diagram, entity-relationship diagram, etc.).